\documentstyle[psfig]{mn}
\def\cm2{cm$^{-2}$}

\newcommand{\lsim}{\ \raise -
2.truept\hbox{\rlap{\hbox{$\sim$}}\raise5.truept
	\hbox{$<$}\ }}			
\newcommand{\gsim}{\ \raise -
2.truept\hbox{\rlap{\hbox{$\sim$}}\raise5.truept
 	\hbox{$>$}\ }}

\begin{document}
\title[High z evolution of optically and IR--selected galaxies]{High
redshift evolution of optically and IR--selected galaxies: a
comparison with CDM scenarios }

\author[A. Fontana et al.]
{
Adriano Fontana$^1$, Nicola Menci$^1$,  Sandro
D'Odorico$^2$, Emanuele Giallongo$^1$,\cr
Francesco Poli$^1$, Stefano Cristiani$^{3,4}$ , Alan Moorwood$^2$, 
Paolo Saracco$^5$\\
$^{1}$ Osservatorio Astronomico di Roma, 00040 Monte Porzio, Italy \\
$^{2}$ European Southern Observatory, 85748 Garching bei M\"unchen, Germany\\
$^{3}$ Dipartimento di Astronomia dell' Universit\`a, 35122
Padova, Italy\\
$^{4}$ Space Telescope European Coordinating Facility, European Southern Observatory, 85748 Garching bei M\"unchen, Germany\\
$^{5}$ Osservatorio Astronomico di Brera,
22055 Merate (LC) Italy}

\maketitle
\begin{abstract}
A combination of ground--based (NTT and VLT) and HST (HDF--N and
HDF--S) public imaging surveys have been used to collect a sample of
1712 I--selected and 319 $K\leq 21$ galaxies observed with an extended
spectral coverage from the U to the K band. Photometric redshifts have
been obtained for all these galaxies, using a spectral library
computed from Bruzual and Charlot models. The results have been
compared with the prediction of an analytic rendition of the current
CDM hierarchical models for galaxy formation that explicitly accounts
for magnitude limits and dust extinction.  We focus in particular on
two observed quantities: the galaxy redshift distribution at $K\leq
21$ and the evolution of the UV luminosity density.  The former has
been proposed by Kauffmann and Charlot (1998) as a very robust
prediction of any CDM hierarchical model, and we show that it is
remarkably constant among different cosmological models. The derived
photometric redshift distribution is in agreement with the
hierarchical CDM prediction, with a fraction of only $5\%$ of galaxies
detected at $z\geq 2$. This result strongly supports
hierarchical scenarios where present--day massive galaxies are
the result of merging processes.  The observed UV luminosity density
in our I-selected sample is confined within a factor of 4 over the
whole range $0 < z < 4.5$, in agreement with previous spectroscopic
and photometric surveys. CDM models in a critical ($\Omega =
1$,$\Lambda =0$) Universe are not able to produce the density of UV
photons that is observed at $z\geq 3$. CDM models in
$\Lambda$--dominated universe are in better agreement at $3 \leq z
\leq 4.5$, but predict a pronounced peak at $z\simeq 1.5$ and a drop
by a factor of 8 from $z=1.5$ to $z=4$ that is not observed in the
data.  We conclude that improvements are required in the treatment of
the physical processes directly related to the SFR, e.g. the starbust
activity in merger processes and/or different recipes for linking 
the supernovae feedback to the star
formation activity.

\end{abstract}
\nokeywords

\section{Introduction}
Understanding how massive galaxies formed and evolved is one of the
major goals of present--day cosmology.  Currently favoured theoretical
scenarios attempt to describe ``ab initio'' the global formation and
evolution of galaxies from primordial fluctuations including the main
physical processes involved (e.g. Kauffmann, White \& Guiderdoni 1993,
Cole et al 1994, Baugh et al 1998). These ``hierarchical'' models
naturally predict galaxies to form from smaller units that accrete gas
and merge to build up present--day massive objects.  These models are
challenged by several observations suggesting that bulges and
ellipticals were formed at a very early stage of the Universe and
slowly evolved thereafter (Bernardi et al 1999, Schade et al 1999 and
references therein).

Kauffmann and Charlot (1998; KC98 hereafter) proposed the use of the
redshift distribution of K--band limited samples to address this
issue.  The main advantage here is that the K band traces the IR
radiation produced by ordinary stars at any $z\leq4$ (see Fig. 1 of
KC98) and is little affected by dust extinction , and it is therefore
a reliable tracer of the mass in stars already assembled in galaxies
at any redshift.  As shown by KC98,  hierarchical models prevent
massive galaxies to be already assembled at $z\geq1$, and the expected
number of galaxies at $z\geq 1.5$ is about 4 times lower than the
 predictions of Pure Luminosity Evolutionary (PLE) models.
Unfortunately, the existing spectroscopic surveys (Cowie et al. 1996,
Cohen et al. 1999) still lack the required completeness at  faint
IR magnitudes in this critical redshift regime.

The evolution of the global star--formation rate as a function of $z$
has long been recognized as a powerful tool to trace galaxy evolution.
First results from spectroscopic (Lilly et al. 1995) and
color--estimated (Madau et al 1996, 1998, Connolly et al 1997)
redshift surveys suggested a steep rise and fall of the SFR with a
main peak at $z\simeq 2$. Photometric redshift analysis (Giallongo et
al 1998, G98 hereafter, Pascarelle et al 1999) and spectroscopic
surveys at low and high $z$ (Treyer et al. 1998, Cowie et al. 1999,
Steidel et al. 1999), inclusion of dust corrections and far--IR
detections (Hughes et al 1998) are now modifying this picture.

In this work we have used deep ground-based and HST multi-band
observations from UV to IR to obtain photometric redshifts for
galaxies in an optical ($I<27.5$) and  IR ($K<21$) sample. The
redshift distributions of the K-band limited sample and of the UV
luminosity density have been compared with the results of the CDM
models to test their fundamentals properties.

\section{The basic ingredients:  Data and models}
\subsection {Multicolor catalogs}

The observations used in this paper cover the full wavelength range
from the UV to the K band and have sub-arcsec image quality.  Two of
the fields were observed with the ESO 3.5 NTT SUSI imager: the first
(hereafter BR1202) is centered on the z=4.7 QSO BR1202-07, the second
is a neighbouring field (NTT Deep Field, NTTDF hereafter). The BVRI
images and catalogs of these fields are described in G98 and Arnouts et al (1999a) respectively, complemented by
NTT observations in J and K (Saracco et al. 1999) and in the U
band. The latter observations and the procedures to obtain the final
UBVRIJK catalogs are fully described in Fontana et al. (in
preparation).

The third dataset results from the VLT-NICMOS observations of the
HDF-S (Fontana et al 1999).
 
Finally, we have used the HDF-N and HDF--S with the IR observations
obtained at Kitt Peak and at NTT--SOFI (Da Costa et al, 1998),
respectively.  For the HDF--N we have used the multicolor catalog
published by Fernandez--Soto et al (1999), which uses an optimal
technique to match the optical and IR images that have a quite
different seeing. A similar catalog for the HDF--S has been provided
by the same authors and is available at the WEB address {\sf
http://www.ess.sunysb.edu/astro/hfds/home.html}.  Only WFPC bands have
been used in the optical, for consistency with the HDF--N.  In both
cases we have clipped the outer regions of the frame with lower S/N.

Despite different origins, these data are sufficiently homogeneous
for the purpose of the paper.  Indeed, all catalogs have been obtained
with similar procedures and software, and we have obtained photometric
redshifts only for objects that are significantly above the detection
threshold, so that small differences in the detection procedures are
not expected to be important.  All the multicolor catalogs use the
optical images as detection frame, as is appropriate for the estimate of
the UV luminosity density at high redshift. We have also performed an
independent object detection on the K images alone to ensure that all the
galaxy at $K<21$ were included in our optically-selected catalogs.

\subsection{Photometric redshifts}

The multicolor catalogs have been used to derive photometric redshifts
for all the galaxies in the sample, using a code already described
elsewhere (G98, Fontana et al in preparation).  The code is based on the
synthetic models of the Bruzual and Charlot GISSEL library, with the
addition of intergalactic absorption (Madau et al 1996) and dust
reddening (SMC--like Pei 1992). The accuracy on the HDF--N
spectroscopic sample is $\sigma _z\sim 0.06 (0.3)$ in the redshift
interval $z=0-1.5 (1.5-3.5)$.

At fainter flux levels the reliability of photometric redshifts has
been estimated with Monte Carlo simulations (Arnouts et al 1999b).  We
have defined a bright sample at  $I_{AB}\le26$ that
includes a subsample of the two HDFs and the NTTDF, and a fainter
HDF(N+S) sample to $I_{AB}\le27.5$. 

It is known that Galactic stars are significant sources of false
high redshift candidates, especially in the brightest samples (Steidel
et al 1999). Obvious stars have been excluded at $I_{AB}\leq 25.5$ in
the HDF-S on the basis of the SExtractor morphological classification
(Arnouts et al 1999a). The morphological selection removes all the
$z\geq5$ candidates in the HDF--S. These objects are typically bright
($I = 20 - 24$) and are always detected in the JHK bands. They would
be formally assigned to $z\geq 5$ since they are nearly undetected in
V and have a negative $J-K$, typical of M star spectra.
Analogously, we have used the detailed morphological classification
developed in Poli et al (1999) to identify stars down to $I_{AB} \simeq
25.3$ in the NTTDF sample.

\subsection{An analytical rendition of hierarchical models}
To compare these results with the present understanding of galaxy
formation and evolution we have developed an analytical rendition of
the hierarchical models (e.g. Cole et al. 1994). The prescription used
to treat all the physical processes involved are identical to the
"Durham" rendition, and we refer the reader to their list of papers
for the details, while the complete formulations of our analytical
rendition are given in Poli et al 1999 (see also Menci and Cavaliere
1999). Rather than following the history of each halo within a Monte
Carlo scheme, we produce the statistical distributions of the main
physical properties of galaxies in the DM halos. Our treatment extends
the White and Frenk 1991 approach explicitly including  the merging of
galaxies in common halos through dynamical friction. This is
accomplished by computing for all galaxies in DM halos the probability
that the dynamical friction time is smaller than the halo survival
time (as given by Lacey and Cole 1993).  For the average quantities of
interest here, this approach produces the same outputs as the
``Durham'' approach.  We emphasize that we have not attempted to
modify or improve the current models, that have several free
parameters tuned to match the local properties of galaxies (counts,
the I-band Tully-Fisher relation and the B-band luminosity function),
since our aim is simply to compare their prediction with the picture
emerging from our data.

The only improvement introduced is a self--consistent treatment of the
dust absorption, by defining an effective optical depth $
\tau_\lambda$(Guiderdoni \& Rocca--Volmerange 1987) that is used to
suppress the expected luminosity:
\begin{equation}
\tau_\lambda = \tau_{dust}^0 (1-\omega)^{1/2} (A_\lambda/ A_V) (Z_g/ Z_\odot)^s f_g
\end{equation}

where $ (1-\omega)^{1/2} (A_\lambda/ A_V) (Z_g/ Z_\odot)^s$ is 
a metal--dependent extinction law, 
$f_g$ is the gas fraction $f_g = m_g / (m_g+m_*)$ (computed
by the code) and $\tau_{dust}^0 $ is
a gas--to--dust ratio chosen to match the observed
B luminosity function at low redshift (Somerville and Primack 1998).

We adopt in this paper four different models, a Standard CDM
(SCDM, $\Omega=1, \Lambda = 0, h=0.5$ ), an Open CDM ($\Omega=0.5,
\Lambda = 0, h=0.7$), a low-density flat model ($\Lambda$CDM,
$\Omega=0.3, \Lambda = 0.7, h=0.6$) and a tilted model, ( $\Omega=1,
\Lambda = 0, h=0.5$). The power spectrum normalization and the 
parameters that describe the IMF, the star formation process and the 
galaxy merging  are taken from Heyl
et al. (1995) for SCDM, Open and $\Lambda$CDM models,
 and  from Poli et al (1999) for the tilted model.

\section {The K$<21$ samples}  
  
\begin{figure}	  
 \begin{minipage}{85mm}
\centerline{\hspace{0cm}\psfig{figure=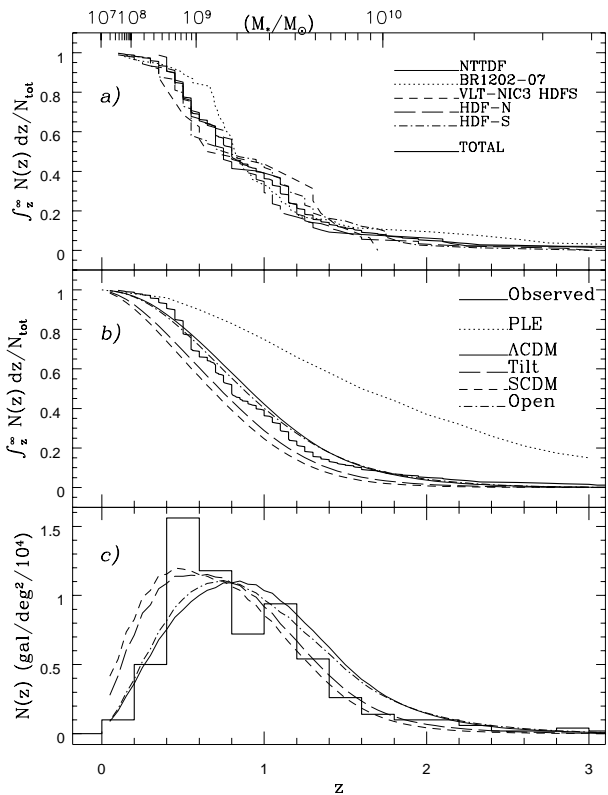,width=0.99\textwidth,angle=0}}
\caption{Redshift distribution at $K\leq 21$.  {\it a)} Observed
normalized cumulative distribution of a total of 391 galaxies in five
fields. Thin lines: individual distributions (see legend for
details). Thick line: total distribution.  {\it b)} Comparison between
the total cumulative distribution of the upper panel with
theoretical predictions of the four CDM models described in the text
and of the PLE model. {\it c)} Comparison between the observed
differential redshift distribution and the CDM
predictions. The upper axis of panel {\it a} shows
the mass in stars contained in an
evolved galaxy at the corresponding redshift, normalized to K=21. A
 Miller--Scalo IMF, age of 2Gyr, solar metallicity and
star--formation timescale of 0.1 Gyrs were adopted. 
}
\end{minipage}
\label{nzK}
\end{figure}  
 
\begin{figure*}	 
\begin{minipage}{140mm}
\centerline{\hspace{0cm}\psfig{figure=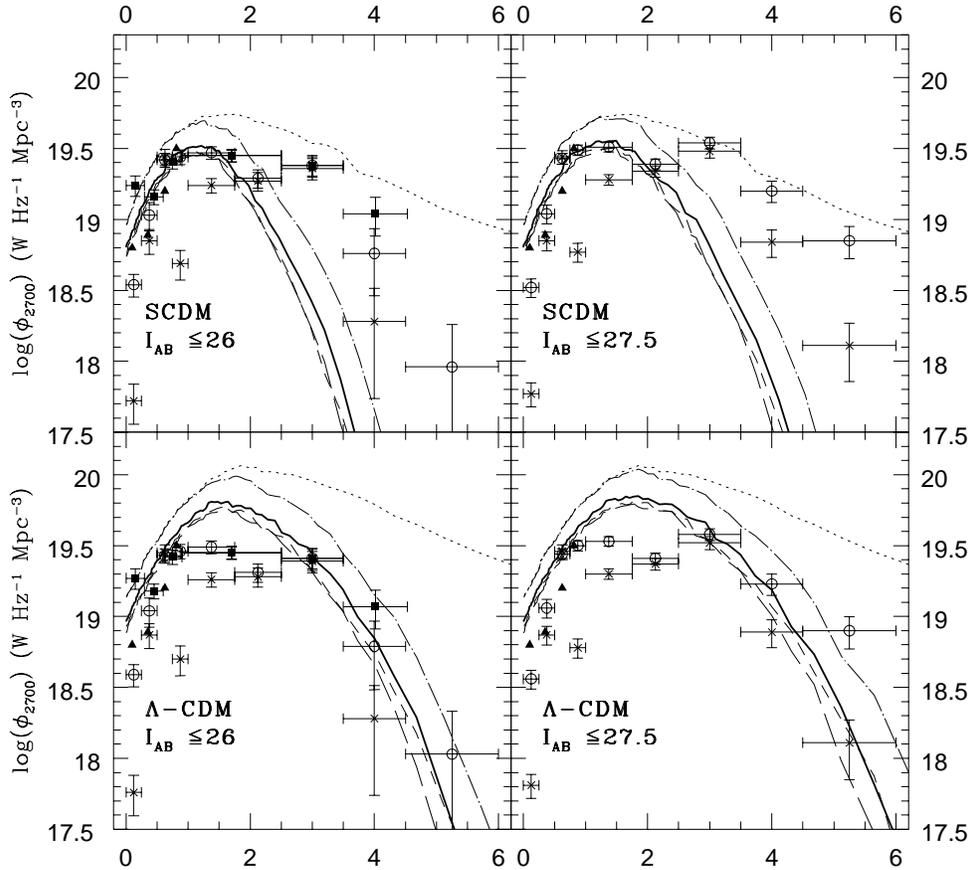,width=0.99\textwidth,angle=0}}
\caption{A comparison between the observed evolution of the luminosity
density $\phi_{2700}$ and the predictions of hierachical models. Empty
circles are derived from the HDF--N, crosses from the HDF--S and
filled squares from the NTTDF. Points are {\it not} corrected for
incompleteness. Errorbars are computed from the number of objects in the
bins, assuming simple Poisson statistics. 
Triangles are taken from the spectroscopic surveys of
Treyer et al; 1997 (lowest bin) and Lilly 1995.  From left to right
two different magnitude limits have been adopted as shown in the
figures.  Upper panels are for standard CDM, lower panels for a
$\Lambda$ dominated Universe. Theoretical curves are computed with our
analythical rendition of hierarchical models for the relevant
cosmology. Dotted line is the total UV luminosity density computed
from the models assuming no dust absorption;  dashed--dot line is the
same quantity when a magnitude cut corresponding to the magnitude
limit in the data is applied; thick solid: magnitude cut + dust
absorption (Calzetti); short dashed: magnitude cut + dust absorption
(SMC) ; long dashed: magnitude cut + dust absorption (MW) }
\end{minipage}
\label{epsi2700}
\end{figure*} 

The normalized cumulative redshift distributions of the 319 objects in
the K--limited sample is shown in Fig.1. The upper panel shows
the different distributions in the five fields considered here as well
as the total distribution.  The results from the five fields are
clearly consistent, within the observed field--to--field scatter. As
expected, the most discordant distributions come from the two smallest
fields (BR1202 and VLTTC).  For the purpose of the KC98 test, what is
critical is the number of massive high~$z$ galaxies detected in the
sample. Only 9\% of the galaxies are found at $z\geq 1.6$ (8\%, 7\%
and 12\% in the NTTDF, HDF--N and HDF--S respectively), and only
5\% at $z\geq\ 2$ (6\%, 4\% and 2\%). A similar conclusion was reached
by Saracco et al 1999.

In the lower panel, the total distribution is compared to the
predictions of PLE models (as adopted from KC98) and of the CDM models
described in the previous section.  All the CDM models are reasonably
consistent with each other, despite the wide variety of cosmological
and physical parameters adopted. As expected, the Open and the
$\Lambda$ models predict the slower evolution. Conversely, there is a
large difference between the ensemble of hierachical models and the
PLE predictions in the KC98 rendition.  This fact, that was already
stressed by KC98, applies not only to the SCDM used by KC98 but also
to other models with different cosmologies, strenghtening the validity
of the KC98 test.

A main result of this paper is that {\it the observed cumulative
distribution is in good agreement with that predicted by the
hierarchical models}, strongly supporting hierarchical scenarios where
massive objects are the result of merging processes in recent epochs.

To reiterate that the $K \leq 21$ threshold corresponds to a
selection with respect to the mass of the galaxies we have labelled the
upper x-axis of Fig.1 with the mass in stars contained in an
evolved galaxy at the corresponding redshift, normalized to K=21. 
At $z\geq 1.6$, the minimum
mass in stars in objects at $K\leq 21$ is  $\simeq
10^{10}M_{\odot}$.  However, a significant contribution to the K
luminosity may also be due to the AGB population during a starburst
phase. For comparison, the same K=21 luminosity may be obtained from a
$z=1.6$ galaxy of only 0.1 Gyrs of age with a constant star formation
rate of $10 M_{\odot}/$yr (and hence a mass of $10^{9}M_{\odot}$).  In
conclusion, both massive evolved objects and/or strong starbursts may
contribute to the counts at $z\geq 1.6$ and $K\leq 21$. Since both
classes are relatively rare in ``bottom--up'' hierarchical models, the
redshift distribution is a sensitive test of these models.

The agreement between the observed and predicted distribution is
slightly worse at $z\leq 0.5$.  The observed {\it differential}
distribution (Fig.1{\it c}) has indeed a paucity of galaxies at very
low redshifts with respect to the theoretical predictions, an effect that
produces the steeper cumulative distribution shown in Fig.1{\it b}.  This is
likely to be a combination of selection effects (these small fields have
been explicitly chosen to be free of bright local galaxies) and of
the slope of the faint end of the  luminosty functions in CDM models,
that is steeper than locally observed.

\section{The cosmological evolution of the UV luminosity density}

We show in Fig.2 the cosmological evolution of the UV
luminosity density $\phi_{2700}$ as estimated from the photometric
redshifts at $I_{AB}\leq 26$ (HDF--N + HDF--S +NTTDF, left panels) and
at $I_{AB}\leq 27.5$ (HDF--N + HDF--S, right panels), for two
different cosmologies. The $L_{2700}$ luminosity of each galaxy in the
sample is directly obtained from the best--fitting spectrum, and falls
in the range of the observed magnitudes at any redshift $z>0.25$.  At
$z>2.4$, most objects are undetected in the IR bands, and the fitting
spectra are constrained by the corresponding upper limits.

The availability of these three different fields allows us for the first
time to compare the evolution of $\phi_{2700}$   in different fields.
At $I_{AB} \le 26$ the NTTDF is in good agreement with  the HDF--N, while the
HDF--S shows significant differences at $0.75 < z < 1.5$ and at $z>3.5$,
due to the variance in the total counts and
in the redshift distributions.
At this stage, it is not clear whether the overall
discrepancy among the fields, and most notably between
HDF--N and HDF--S, is due to a real cosmic variance or to some
instrumental effect, and we consider it as an estimate of the global
uncertainities in this analysis.
At $z\leq 1.5$, the
results from the NTTDF and the HDF--N are consistent with
those from spectroscopic surveys (Treyer 1998, Lilly 1995) when the
corrections for steep luminosity functions are adopted, as seems
appropriate for fields dominated by blue star--forming galaxies.  At
higher $z$ all the fields concur to a scenario where the UV luminosity density
is relatively constant from $z=1$ to $z=4.5$.

The overall picture emerging from Fig.2 is that the UV
luminosity density does not change by more than a factor of 4 over the
redshift range $0 < z < 4.5$, the only exception beeing the $z\simeq
4$ redshift bin in the HDF--S (but see below). 

The comparison with CDM models is less straightforward here.  We have
overplotted in Fig.2 the prediction of two well--studied examples, the
SCDM and $\Lambda$--CDM. At variance with previous works, we have not
corrected the observed values for incompleteness or extinction, but
rather we have explicitly shown the differential effects of the
inclusion of a magnitude limit and different dust extinction curves on
the theoretical expectations. The comparison shows that the
$\phi_{2700}$ overall shape is hardly recovered by the current CDM
models. In particular, the SCDM model is not able to produce the
density of UV photons that is observed at $z\geq 3$, while
$\Lambda$--CDM is in better agreement at $3 \leq z \leq 4.5$, but
predicts a pronounced peak at $z\simeq 1.5$ and a drop by a factor 
of 8 from $z=1.5$ to $z=4$ that is not observed in the data.

A more accurate comparison at high redshift can be carried out by
plotting (Fig.3) the redshift evolution of the UV
luminosity density at a shorter wavelength (1400~\AA), where the
best--fitting spectra are tied to the observed R and I bands, and
comparing it with spectroscopic surveys and CDM models.  Photometric
surveys are consistent with the results of spectroscopic surveys on
brighter samples with the exception of the HDF--S, especially at
$z\geq 3.5$.  The large variance between the HDF--N and the HDF--S is
due to an intrinsic lack of high redshift galaxies in the latter.  In
particular 4 objects are identified at $z\geq 5$ in the HDF--N, while
no convincing candidate is found in the HDF--S, after removing obvious
stars.  It should be noted that when a standard color selection as in
Madau et al. (1996) is applied, a comparable number of B-dropout
galaxies can be found in the HDF-N and -S. However, these represent
only a fraction of the high redshift galaxies found by the photometric
redshift technique (see Pascarelle et al. 1998, Fontana et al in
preparation) that uses the IR bands as additional constraints.  These
additional high-z candidates are brighter and more numerous in the
HDF--N than in the HDF--S, producing the different values shown in
Fig. 3.

\begin{figure}
 \begin{minipage}{85mm}
\centerline{\hspace{0cm}\psfig{figure=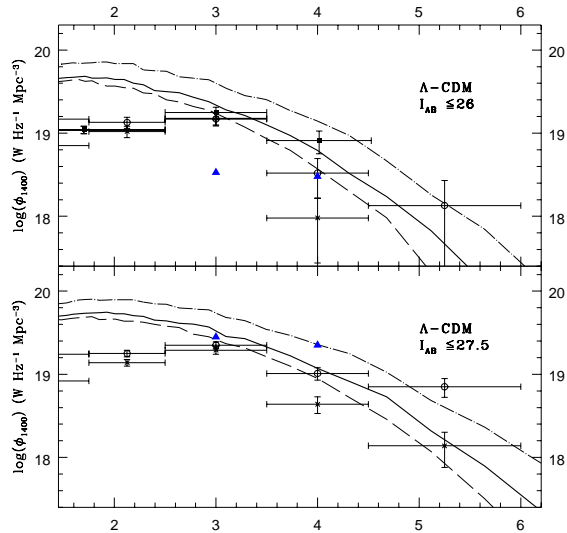,width=0.99\textwidth,angle=0}}
\caption{A comparison between the observed evolution of the
luminosity density $\phi_{1400}$ and the prediction of hierachical
models.  Triangles in the upper panel are from spectroscopic data of
Steidel et al. (1999). These have been corrected for incompleteness up
to $I_{AB}=27.5$ in the lower panel. All the other symbols and lines
as in Fig. 2 }
\end{minipage}
\label{epsi1400}
\end{figure}

\section {Summary}

We have collected and analyzed a sample of 1712 I--selected and  319
$K\leq21$ galaxies from public deep imaging surveys, mainly the two
HDFs and the NTTDF. We have derived photometric redshifts for the
whole sample in an homogeneuos way.
The results may be summarized as follows:
   
\begin{itemize}

\item The redshift distribution of the $K\leq 21$ sample (Fig. 1) is
dominated by objects at low or intermediate redshift, with a fraction
of only $9\%$ of galaxies detected at $z\geq 1.6$ and of $5\%$ at
$z\geq 2$;

\item The UV luminosity density $\phi_{2700}$ is confined within a
factor of  4 from $z=0$ to $z=4.5$ (Fig.2), and is
relatively constant from $z=1$ to $z=4.5$, although significant
field--to--field variations exist and dominate over statistical
uncertainties;

\item A comparison between HDF--N and HDF--S shows that the UV
luminosity density $\phi_{1400}$ at $z\geq 4.5$ 
is still poorly determined, probably
due to the cosmic variance 
between these two small fields. $\phi_{1400}$ changes by a factor of
$\simeq 6$ between the two fields at $z\geq 4.5$, since no convincing $z\geq
5$ candidate 
is found in the HDF--S, compared to 4 in the HDF--N (2 of which have
spectroscopic confirmation).

\end{itemize}

We have compared these results with the predictions of our analytical
rendition of popular CDM models. We chose not to correct the data for
incompleteness or dust extinction, but rather to include both effects
in the theoretical model we compare with. 

The $K\leq21$ redshift distribution at $z\geq 1$ directly reflects the
number of massive galaxies already assembled at $z\geq1$ (KC98).  The
agreement that we find between the observed distribution and the
prediction of an ensemble of CDM models (Fig.1{\it b}) strongly supports
a key feature of these
 theoretical scenarios, i.e. that   massive objects are the result of
merging processes in recent epochs.

On the other hand the overall shape of the UV luminosity density,
that is tied to the physical mechanisms driving the star formation
processes, is not easily reproduced by current CDM models. The
comparison between the observed evolution and the prediction of two
different models (SCDM and $\Lambda$--CDM) shows that the SCDM model is
not able to produce the density of UV photons that is observed at
$z\geq 3$.
Given the $I_{AB}\leq27.5$ limit applied to both observations and models,
the discrepancy means that the current SCDM
model fails to reproduce the bright tail of the luminosity function.
$\Lambda$--CDM is in better agreement at $3 \leq z \leq 4.5$, but
predicts a pronounced peak at $z\simeq 1.5$ and a drop by a factor 
of 8 from $z=1.5$ to $z=4$ that is not observed in the data. Such a
result holds for all the adopted extinction laws. This implies that
further refinements are required in the treatment of the physical
processes directly related to the SFR.  For instance, adopting a
weaker feedback would increase the luminosity of fainter galaxies that
dominate the statistics at $z\geq 2$ yielding a less steep decline of
the SFR.  Another possibility is that merging activity at $z\geq 2 $
is effective in enhancing the luminosity and/or the number density of
faint galaxies at such $z$. A first attempt to include these effects has been
described by Somerville, Primack and Faber 1998.
These - or other - changes will require a
global recalibration of the model parameters, in order to fit the
increasing number of observables at low and high redshift.

{\bf Acknowledgments}\\

We thank Alvio Renzini for stimulating discussions on this research
topics, the referee, Guinevere Kauffmann, for several useful comments
which improved the paper and F. Governato, S. Savaglio and V. Testa
for comments on earlier versions of this work.  The paper is based on
observations made with: the ESO VLT Antu telescope at the Paranal
Observatory, the ESO New Technology Telescope at the La Silla
Observatory (some of which under the EIS programs 59.A-9005(A),
60.A-9005(A)), the NASA/ESA Hubble Space Telescope and the Kitt Peak
National Observatory.  The ultraviolet observations of the NTTDF were
performed in SUSI-2 guaranteed time of the Observatory of Rome in the
framework of the ESO-Rome Observatory agreement for this instrument.

\end{document}